\documentclass[fleqn,10pt]{wlscirep}
\usepackage{float}
\title{Demonstration of quantum synchronization based on second-order quantum coherence of entangled photons}

\author[1,2]{Runai Quan}
\author[1,2]{Yiwei Zhai}
\author[1,2]{Mengmeng Wang}
\author[1,2]{Feiyan Hou}
\author[1,2]{Shaofeng Wang}
\author[1,2]{Xiao Xiang}
\author[1]{Tao Liu}
\author[1]{Shougang Zhang}
\author[1,*]{Ruifang Dong}
\affil[1]{Key Laboratory of Time and Frequency Primary Standards, National Time Service Center, Chinese Academy of Sciences, Xi'an, 710600, China}
\affil[2]{University of Chinese Academy of Sciences, Beijing, 100049, China}

\affil[*]{dongruifang@ntsc.ac.cn}

\begin{abstract}
Based on the second-order quantum interference between frequency entangled photons that are generated by parametric down conversion, a quantum strategic algorithm for synchronizing two spatially separated clocks has been recently presented. In the reference frame of a Hong-Ou-Mandel (HOM) interferometer, photon correlations are used to define simultaneous events. Once the HOM interferometer is balanced by use of an adjustable optical delay in one arm, arrival times of simultaneously generated photons are recorded by each clock. The clock offset is determined by correlation measurement of the recorded arrival times. Utilizing this algorithm, we demonstrate a proof-of-principle experiment for synchronizing two clocks separated by 4km fiber link. A minimum timing stability of 0.4 ps at averaging time of 16000 s is achieved with an absolute time accuracy of 59.4 ps. The timing stability is verified to be limited by the correlation measurement device and ideally can be better than 10 fs. Such results shine a light to the application of quantum clock synchronization in the real high-accuracy timing system.
\end{abstract}

\begin{document}

\maketitle
\thispagestyle{empty}

\section*{Introduction}

Accompanying with the remarkable improvements in the ability of generating and measuring high-accuracy time-frequency signal\cite{Bloom2014,Huntemann2012,Hinkley2013}, the application of high-precision time and frequency is playing an increasingly important role in areas of fundamental physics (e.g., tests of fundamental physical constants\cite{Peik2004,Fischer2004,Fortier2007,Lea2007}, general relativity verification\cite{Muller2007,Chou2010,Reynaud2009}, dark matter searching\cite{Derevianko2014}), and many other areas of highly advanced technology and engineering infrastructure (such as long-baseline interferometry for radio astronomy\cite{Shillue2004}, accelerator-based x-ray sources\cite{ssrl,Decamp2001,Schoenlein1996,Cavalieri2005}, mapping of the Earth's geoid\cite{Bondarescu2012}, deep space exploration\cite{Krisher1990,Calhoun2002,Calhoun2004}, precise distance ranging and timing\cite{Ye2004,Coddington2009,Giorgetta2013}, etc). The breakthrough in all these applications will benefit from the capability of high-stability time and frequency transfer, therefore attempts to significantly improve the conventional time and frequency transfer accuracy becomes a crucial subject.

Accurate time transfer between two geographically distant sites is currently dominated by satellite-based navigation systems. A traditional method for transferring frequency and time standards over long distances has been the common-view global positioning system (GPS)\cite{Levine1999}. By averaging for about a day it is possible to reach accuracies of one part in 1E-14\cite{Lee2008}. Subsequently, the two way satellite time and frequency transfer (TWSTFT)\cite{Hanson1989, Kirchner1999, Bauch2009} has pushed the frequency transfer instability to the low parts in E-15 in 1 day. With a much higher frequency and bandwidth of laser pulses than radio radiations, the time transfer by laser link (T2L2) was proposed and recently shown an synchronization accuracy of tens of picoseconds\cite{Guillemot2008, Laas-Bourez2015}, which corresponds to a frequency stability of E-16 in 1 day. However, these techniques are far from enough to satisfy the growing requirement for comparison of the new generation of high-precision atomic clocks. An alternative for stable transferring of the time and frequency signal is transmission over optical fibers since an environmentally isolated fiber can be considerably more stable than free space paths. Currently, multiple research teams are oriented at ultrastable frequency transfer through optical fibers
\cite{Lopez2008,Marra2011,Predehl2012,Lopez2015}, and a frequency instability of $10^{-20}$ after $10^3$ s averaging time has been demonstrated\cite{Lopez2015}. However, time scale comparisons are always necessary, therefore methods on time transfer through optical fibers(TTTOF) are also under investigation\cite{Ebenhag2010,Ebenhag2008,Piester2011,Rost2012,Smotlacha2010,Wang2012,Lopez2013,Lessing2015}. Among all the achieved results, a timing stability of the time transfer over 158 km long optical link with minimum value of 300 fs in terms of time deviation (TDEV) at averaging time of 10 s has been presented \cite{Lessing2015}. Moreover, it has been demonstrated that the long-term time transfer stability is less sensitive to the fiber length than the method accuracy\cite{Smotlacha2010}.

The accuracy of the above time transfer procedures is classically limited by the available power and bandwidth\cite{Giovannetti2001a}. To overcome the classical limits, quantum clock synchronization (QCS) was proposed. By utilizing the nonclassical and nonlocal characteristics of quantum entangled and squeezing resources, QCS can result in enhanced accuracy compared with their classical analogues. According to the resources being adopted in the system, there are two families of protocols for QCS. The original QCS protocol was proposed to utilize shared prior quantum entanglement, and classical communications, to establish a synchronized pair of atomic clocks\cite{Jozsa2000}. In contrast to classical synchronization schemes, the accuracy of this protocol is independent of knowledge of their relative locations or the properties of the intervening medium. Based on this idea, a general framework\cite{ZhangYL2013} as well as several multiparty clock synchronization protocols\cite{Krco2002,Ben-Av2011} were further proposed. However, since the shared prior entanglement between the distributed quantum clocks is hard to establish, the above QCS protocols have a vital shortcoming. On the other hand, quantum versions of Einstein synchronization protocol was proposed by Giovannetti \emph{et al.}\cite{Giovannetti2001a} in the beginning of this century. By employing frequency entangled pulses, it can break through the shot noise limit on the classical timing system, and finally reaches the fundamental Heisenberg limit. Since then diverse QCS algorithms flourished\cite{Giovannetti2001b,Giovannetti2002a,Bahder2004,Valencia2004,Giovannetti2004,Ho2009,Hou2012,WangJ2015}. However, experimental implementation focusing on the quantum clock synchronization
is few, except that a one-way synchronization of clocks was reported by Valencia group\cite{Valencia2004}, which accomplished an experimental demonstration at 3 km fiber distance with an accuracy of picosecond.

In this paper, we report a proof-of-principle experiment on synchronizing two clocks separated by 4-km fiber link based on HOM quantum interference between two frequency entangled
photons. In contrast to the one-way synchronization protocol, The synchronization algorithm based on a HOM interferometric coherence \cite{Bahder2004} outperforms the one-way synchronization protocol since the timing stability is independent on the geometric distances between clocks. By accurate control of an optical delay line and on the resolution of second order quantum interference exhibited by correlated photons in a HOM interferometer\cite{Hong1987}, a sub-picosecond or even lower timing stability could be achieved. Based on this algorithm, a long-term synchronizing stability of 0.4 picosecond is achieved, which is mainly limited by the performance of the time-arrivals correlation measurement device and can in principles reaches the level of a few femtoseconds. By improving the generation and detection efficiency of the frequency entangled photon pairs in the experimental scheme, it may find important applications in radio astronomy, such as in VLBI.

\section*{Results}

\subsection*{Principle for synchronizing two clocks based on the HOM interferometric quantum coherence}\label{theory}

A brief description of the synchronization algorithm based on the HOM interferometric quantum coherence\cite{Bahder2004} is addressed here for completeness. The sketch of the proposal is shown in Fig.~1, clock B is aimed to synchronize with spatially separated clock A. We assume that both clocks A and B are stable enough during the synchronization procedure, and have the same rate with respect to common coordinate time. Thus the clock synchronization problem is reduced to identifying the time offset between the two spatially separated clocks. In this algorithm, clocks A and B are assumed stationary in the frame of reference in which the entangled source is at rest. The HOM interferometer is spatially co-located with the photon-pair source. Photon pairs are departed and transmitted to clock A and clock B, respectively, and then reflected back toward the HOM interferometer. Photons that are ``coincident at clock A and clock B'' can thus be defined as those that arrive at the HOM interferometer simultaneously. The minimum photon coincidence events of the HOM outputs indicate that the channel paths, from the photon-pair source to the two clocks sites, are balanced.

To achieve and maintain the balance, a controllable optical delay line is inserted in one path. Via a feedback loop applied onto the optical delay line, the entangled photon pairs arrive coincidentally at the HOM interferometer after propagating through different paths. As soon as the entangled photon pairs are ``coincident at clock A and clock B'', the photon arrival-time data at clock A and clock B, $\{\tau _i^A\}$ and $\{\tau _i^B\}$, $i=1,...,N$, are recorded in a time window $T$ with respect to each local clock. The time $T$ is much longer than the expected clock difference between clock A and clock B. The photon arrival-time data, $\{\tau _i^A\}$ and $\{\tau _i^B\}$ can be assembled into functions ${f_A}(t)$ and ${f_B}(t)$:
\begin{eqnarray}\label{arriv}
 && {f_A}(t) = \frac{1}{{\sqrt N }}\sum\limits_{i = 1}^N {\delta (t - \tau _i^A)} \\\nonumber
 && {f_B}(t) = \frac{1}{{\sqrt N }}\sum\limits_{j = 1}^N {\delta (t - \tau _j^B)}
\end{eqnarray}
The cross correlation between the two functions can be executed to extract the time offset:
\begin{equation}\label{corre1}
 g(\tau) = \int_{ - \infty }^{ + \infty } {{f_A}} (t){f_B}(t - \tau )dt\
\end{equation}
Substituting the timing sets in Eq.~(\ref{arriv}) into Eq.~(\ref{corre1}), and integrating over the whole timing events, we can get the cross correlation in discrete pattern:
\begin{equation}\label{corre2}
 g(\tau) = \frac{1}{N}\sum\limits_{i = 1}^N {\sum\limits_{j = 1}^N {\delta (} \tau  - \tau _i^A}  + \tau _j^B)
\end{equation}
The time offset can be simply found by $\tau=\tau _i^A-\tau _i^B$, since the terms of $i \ne j$ contribute few to the correlation. By adding the time offset $\tau$ to clock B, it is synchronized to clock A.
\begin{figure}[!ht] 
\centering
\includegraphics[width=10cm]{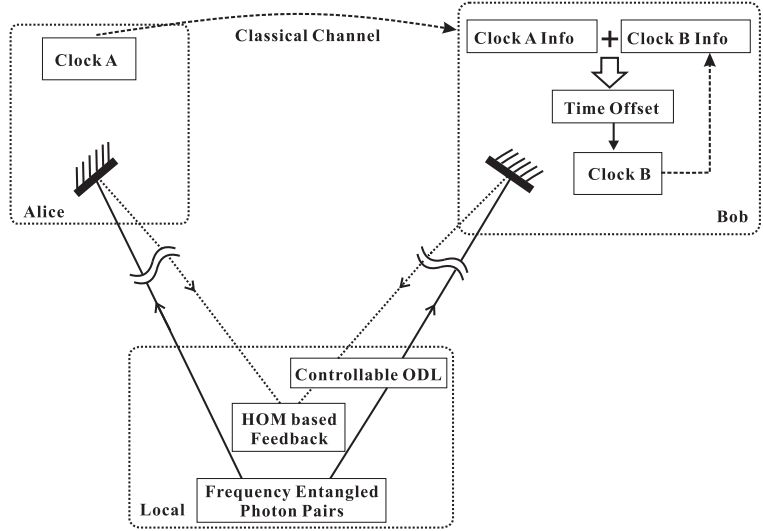}
\caption{The fundamental diagram of quantum clock synchronization system based on the HOM interferometric quantum coherence.}
\label{diagram}
\end{figure}
\subsection*{Experimental realization of quantum time synchronization}\label{experiment}

The proof-of-principle experiment of the above quantum time synchronisation proposal is implemented in fiber channels and the scheme is sketched in Fig.~2. The source of frequency entangled photon-pairs (denoted as signal and idler) with orthogonal polarizations is generated based on a pulse pumped spontaneous parametric down conversion (SPDC) process\cite{Zhang2013,Quan2015}.
Two optical circulators (OCA and OCB) are inserted into the signal and idler arms, respectively. The frequency entangled signal and idler photons are sent into two commercial 2km-long, spooled single-mode optical fiber channels via the transmission from 1st ports to 2nd ports of the individual optical circulators. An optical delay line (ODL,General Photonics Inc.) with a fixed delay value of $\tau_{ODL}$ = 100 ps is inserted in the signal path, and a motorized optical delay line (MDL-002, General Photonics Inc.), which is adjustable from 0 to 560 ps with a resolution of 1 fs, is inserted into the idler path. At the end of each fiber path, the transmitted photons are split into two portions by a 90/10 standard single mode fiber coupler. The 90\% outputs are reflected by faraday rotators and backtracked to the 2nd ports of the optical circulators. From ports 3 of OCA and OCB, the returned photon pairs are fed into a fiber-based HOM interferometer for balancing the two fiber paths and the minimum coincidence counts of the HOM interferometer indicates the balance of the two fiber paths. The remaining 10\% outputs are sent into a pair of fiber-coupled InGaAs single-photon counters (idQuantique id210-SMF-STD-100MHz) D3 and D4, with the external triggering signal extracted from repetition rate of the pump pulses. The clicks recorded by these detectors can be regarded as timing signals for clock A and clock B. Connecting the outputs of D3 and D4 a the commercial Time-Correlated Single Photon Counting (TCSPC) System (Picoquant PicoHarp 300), the time difference between the two simulated clocks A and B is measured. Once the balance is achieved and maintained by a feedback loop, the measured time difference denotes the time offset that needs to be synchronized.

\begin{figure}[H]
\centering
\includegraphics[width=11cm]{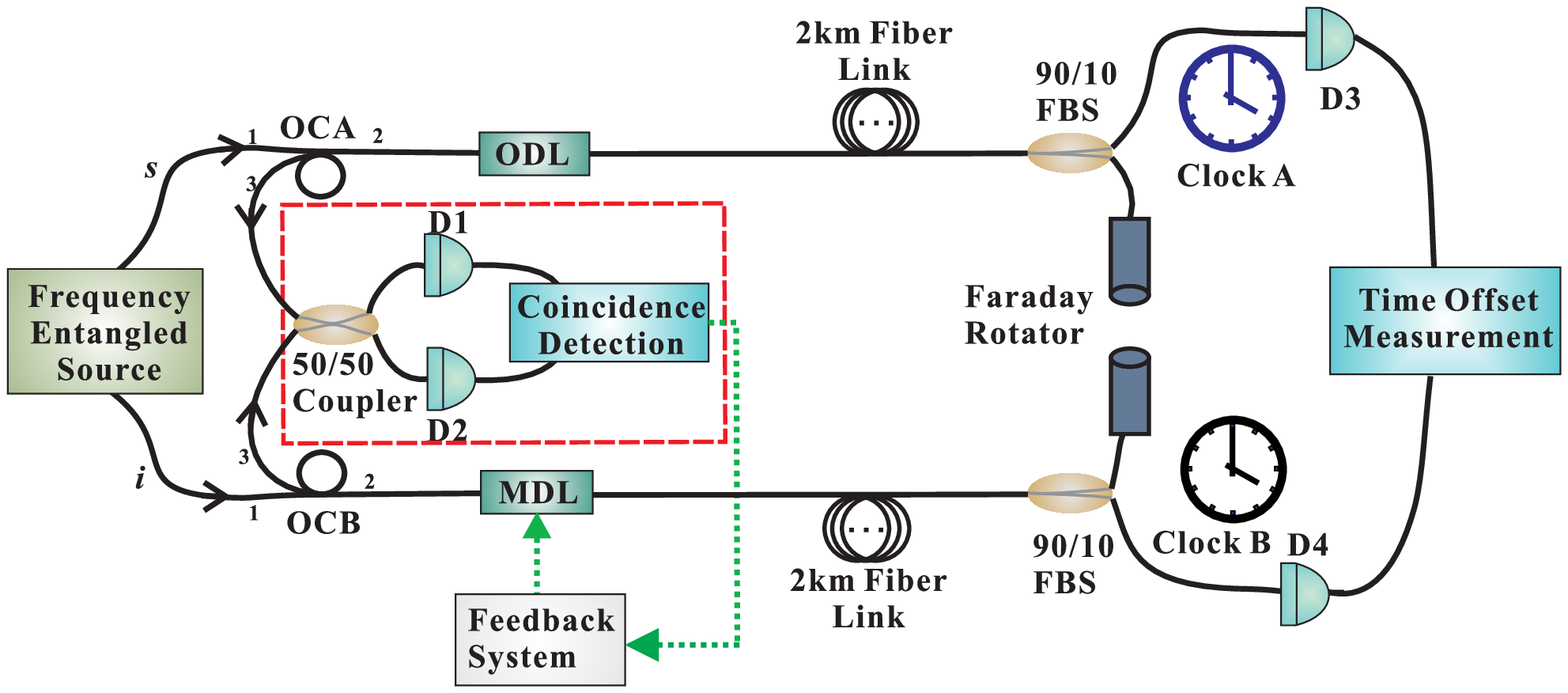}
\caption{Schematic setup of the proof-of-principle clock synchronization experiment based on the HOM interferometer}
\label{QCSsetup}
\end{figure}

\subsection*{Results and analysis}

According to Bahder \textit{et al.}\cite{Bahder2004}, the synchronization stability between A and B is determined by the timing stability of the maintained balance of the fiber paths based on the HOM interferometric quantum coincidence. The delay values of the steered MDL, corresponding to the minimum coincidence of the HOM, are monitored for more than 32 hours and plotted in Fig.~3(a)(black dashed line). By comparing with the simultaneously measured environmental temperature variations (blue solid line), a similar tendency is observed, which shows the dominant effect of temperature fluctuations on the fiber-path fluctuation. While the fiber paths are balanced, the in-loop timing jitter is monitored for evaluating the stability of the maintained balance. The result during a time period of more than 32 hours is plotted in Fig.~3(b). The RMS fluctuation of the in-loop timing jitter is measured to be 0.2 ps. The corresponding time deviation (TDEV) of the jitter versus the averaging time is plotted in Fig.~5(red up-triangles). At averaging time of 1000 s, the in-loop time deviation is below 30 fs; when the averaging time is extended to 10000 s, the time deviation degrades to 8 fs. This TDEV curve sets the lower limit to the timing instability between two remote clocks can be improved to the femtosecond level.
For comparison, the in-loop timing jitter for the case that the two 2km-long, spooled optical fibers are removed is also investigated. The relevant TDEV result is given with blue down-triangles in Fig.~5. It can be seen that, at short term the in-loop TDEV timing jitter after propagating through 4 km of optical fiber paths is half-of-an order of magnitude higher than that of the case when the long-distance fiber is not inserted. However, the long-term timing jitter is independent to the fiber length, which is coincident with the result demonstrated in Ref.\cite{Smotlacha2010}.

\begin{figure}[!ht]
\centering
\includegraphics[width=15cm]{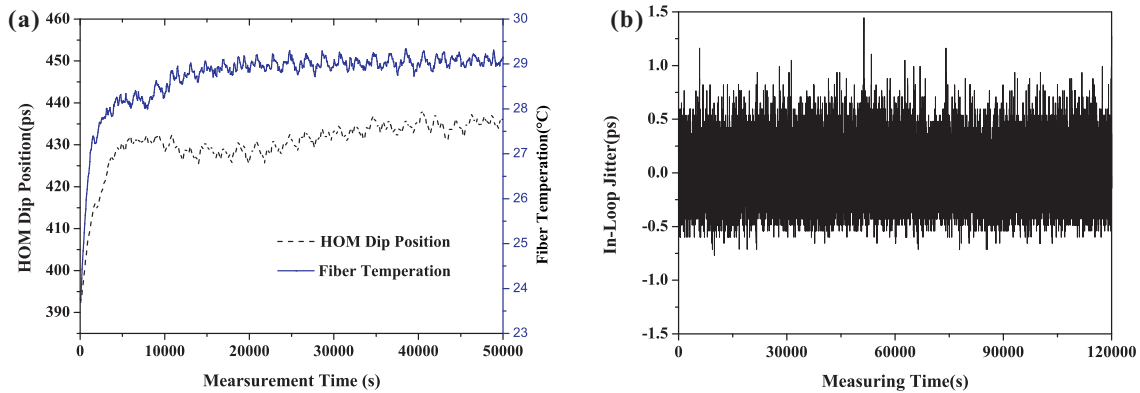}
\caption{(a) The measured MDL movements (black dashed line) and the fiber temperature (blue solid line) versus the recording time. (b) The measured in-loop jitter during a period of more than 32 hours}
\label{feedback}
\end{figure}

Maintaining the feedback loop in operation, the correlation of time arrivals at the two single photon detectors D3 and D4 is measured by a the time arrivals correlation measurement equipment (Picoquant PicoHarp 300). Fig.~4(a) shows the measured correlation distribution of the signal-idler pairs at averaging time of 1000 s for both with (red circles) and without (black squares) 2km fiber links in the setup. Through a Gaussian fitting \cite{Valencia2004}, the time offset $\tau$ is achieved. By successively repeating the correlation measurements, the stability of determining the time offset $\tau$ is evaluated to have a mean value of 868.1ps with a standard deviation of 1.5ps.
The variations of the extracted time offsets in more than 30 hours is displayed in Fig.~4(b). When the 2km-long fibers are removed, the measured time offset is 927.5ps with a standard deviation of 1.47ps. The gap between the two values manifests that such setup for 2km-long fiber propagation has a synchronization accuracy of 59.4 ps.
The TDEV results are displayed in Fig.~5 for both with (red circulares) and without (blue diamonds) 2km fiber links. It can be clearly seen that the timing stability is independent of the geometric distance between two clocks. At averaging time of 1000 s, the measured time offsets give an instability of 1.5 ps. By increasing the averaging time beyond 4000s, the timing instability falls below 1 ps. When the averaging time is larger than 16000s, the instability approaches value of $0.4\pm0.1$ ps.
In order to further illustrate the performance of the stabilized fiber-based links, the time offset fluctuations in terms of TDEV without the HOM interferometer being locked is also given in Fig.~5(olive hexagons). One can see that the time deviation of the free-running links is increased with the averaging time, therefore sub-picosecond stability can never be achieved without the HOM interferometer being locked.

\begin{figure}[!ht]
\centering
\includegraphics[width=15cm]{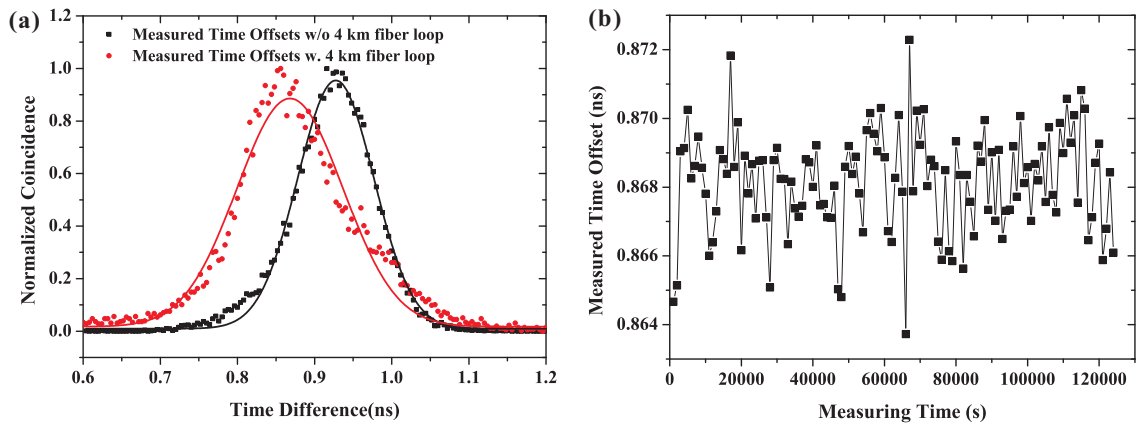}
\caption{(a)The measured timing correlation distributions between two single photon detectors at an averaging time of 1000 s for both with (red circles) and without (black squares) 2km fiber links in the setup. (b) The extracted time offset fluctuations in more than 32 hours.}
\label{coincidence}
\end{figure}

The displayed gap between the time deviations of the measured time offsets and the in-loop noise in Fig.~5 reveals that there is another experimental imperfection which dominates the practical synchronizing instability. To determine the systematic stability of the TCSPC measurement device, two standard pulses per second (PPS) signals are derived from one Hydrogen maser (2.83E-15@1000s) and connected to the start and stop input ports of PicoHarp 300. A set of measurements of the time differences between the two PPS signals is implemented and the timing stability in terms of TDEV is given by black squares in Fig.~5. It shows that, the time deviation of the TCSPC system has a value of 0.9 ps at averaging time of 1000 s and $0.3\pm0.15$ ps at averaging time of 16000 s, which manifests its dominant limitation on the achievable timing stability in our experiment.
\begin{figure}[H]
\centering
\includegraphics[width=8cm]{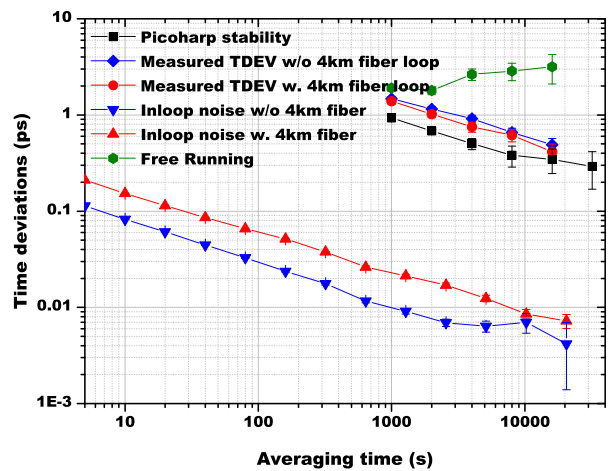}
\caption{The time deviation results for the in-loop timing jitter with(red up-triangles) and without (blue down-triangles) the 2 km fiber links, the systematic instability of PicoHarp 300 (black squares), the measured time offset with (red circles) and without (blue diamonds) 2 km fiber links when the HOM interferometer is locked, and the time offset (olive hexagons) for the free running condition.}
\label{tdev}
\end{figure}

\section*{Discussion}

Based on the second-order HOM quantum interference between frequency entangled photons generated by pulsed SPDC process, we have demonstrated a proof-of-principle experiment for synchronizing two clocks separated by 4-km fiber link. The average time offset between two simulated clocks A and B is measured to be 868.1ps. By measuring the time offset without the two 2km-long fibers in the setup, a value of 927.5ps is achieved. The gap between the two values manifests that such setup for 4km-long fiber path separation has a synchronization accuracy of 59.4 ps. A minimum timing stability of 0.4 ps in terms of time deviation (TDEV) is achieved. Ideally, the synchronization stability is mainly limited by the in-loop jitter of the stabilized system, which can reach several femtoseconds. In the experiment, the synchronizing stability is found to be mainly restricted by the systematic stability of PicoHarp 300, which has a minimum value of $0.3\pm0.15$ ps. This offset corresponds to several systematic imperfections, the fiber length mismatch between the two outputs of the two 90/10 couplers should be among them.

This quantum clock synchronization algorithm based on the HOM second-order quantum interference is in some way analogous to the classical methods on time transfer through optical fibers(TTTOF) since the long-term time transfer stability of both methods is insensitive to the fiber length. Such experimental verification on TTTOF has been demonstrated by Smotlacha \textit{et al.}\cite{Smotlacha2010}. In our case, the limitation of the long fiber distance application is mainly on the flux of the generated signal-idler and the detection efficiency of the single photon detectors, due to the path loss in the fiber channels. If more efficient single photon detectors as well as optimized focusing of pump and collection optics are used, the spectral brightness of the generated frequency entangled photon pairs could be increased by approximately one order of magnitude. Furthermore, the splitting ratio of the 90/10 standard single mode fiber coupler at the end of fiber paths can be played with to shorten the averaging time of the time offset measurement.

\section*{Methods}

\subsection*{Generation of a frequency entangled source}
The frequency entangled photon-pair source is generated based on a pulse pumped spontaneous parametric down conversion (SPDC) process. As shown in the pump source is a commercial Ti:Sapphire laser (Fusion 20-150, FemtoLasers) with a repetition rate of 75 MHz. It produces slightly chirped pulses centered at 789 nm with 3-dB bandwidth of 22 nm. After single passing through a 20 mm, thermally-optimized, type-II PPKTP crystal (Raicol Crystals), a source of frequency entangled photon pairs (denoted as signal and idler) with orthogonal polarizations is generated. For eliminating the residual pump, a series of dichroic mirrors and long-pass filters along the propagation path right after the PPKTP crystal. Afterwards, the photon pairs are collimated and coupled into a fiber polarization beam splitter (FPBS). With the aid of a half wave plate (HWP) in front of the FPBS, the signal and idler are departed and propagate along the two output ports of the FPBS, respectively. By connecting the outputs of the FPBS directly to the single photon detectors (D1 \& D2), single photons count rates of 104k/s and 140k/s are measured respectively, and a coincidence rates of 3k/s is achieved.

\subsection*{Implementation of fiber-based HOM interferometer}
The fiber-based HOM interferometer for measuring the second-order quantum coherence of the photon pairs is based on a 2$\times$2 50/50 fiber coupler. The signal and idler photons propagate from the 1st ports to the 2nd ports of two optical circulators (OCA and OCB), respectively. At the end of each fiber path, the 90\% outputs of the individual 90/10 standard single mode fiber couplers are reflected by faraday rotators and backtracked to the 2nd ports of the optical circulators. From ports 3 of OCA and OCB, the returned photon pairs are connected to the two input ports of the 50/50 coupler. To optimize the interference, A fiber-based polarization controller (FPC) is further applied to accord the polarizations of the signal and idler before interfere them on the 50/50 coupler. The two outputs of the 50/50 coupler were sent to a pair of fiber-coupled InGaAs single-photon detectors (ROI Optoelectronic Technology, SPD4) D1 and D2, which are operated in Geiger mode and externally triggered by 75MHz TTL signal extracted from the repetition rate of pump pulses.

\subsection*{Balancing the fiber paths based on a HOM interferometer}

To balance the signal and idler fiber paths, the measured second-order quantum coincidence of the transmitted signal and idler photons after a HOM interferometer is used to extract the required error signal for feedback loop together with the steering of MDL. According to Ref.[\cite{Quan2015}], the HOM interferogram takes a form of
\begin{eqnarray}
  P_c(\tau) \propto \int \int d{\omega_s} d{\omega_i}(\left|A(\omega_s,\omega_i)\right|^2-\left|A(\omega_s,\omega_i) A(\omega_i,\omega_s)\right|\cos [(\omega_s-\omega_i)\tau]),
\end{eqnarray}
where $\omega_s$ and $\omega_i$ denote the frequencies of the signal and idler photons, respectively. $A({\omega_s},{\omega_i})$ represents the joint amplitude spectral function of the generated photon pairs. $\tau$ represents the relative delay between the signal and idler paths. By scanning the MDL, the HOM interferometric coincidence shows a dip shape with a visibility of 68\% and a coherence time of 3 ps\cite{Quan2015}. Although the measured HOM visibility is reduced due to the nonnegligible effect of the second-order dispersion of PPKTP, it is sufficient for balancing the the fiber paths. The iterative working principle is as follows: First, we implement a complete HOM interferogram measurement and record down the value of MDL corresponding to the minimum coincidence, $\tau_{MDL,0}$. By employing $\tau_{MDL,0}$ as the reference point and applying a square modulation with a depth of $\delta = 0.4$ ps on the motor driver of MDL, two coincidences at $\tau_{MDL,0}-\delta/2$ and $\tau_{MDL,0}+\delta/2$ within 1 s averaging time are measured respectively. We denote them as $R_c(-)$ and $R_c(+)$. By comparing $R_c(-)$ and $R_c(+)$, one can judge whether $\tau_{MDL,0}$ maintains the minimum HOM coincidence. If $R_c(-)>R_c(+)$, it is inferred that $\tau_{MDL,0}$ needs to be decreased. While if $R_c(-)<R_c(+)$, $\tau_{MDL,0}$ needs to be increased. After adding or subtracting an appropriate value of $\Delta$ to the initial set $\tau_{MDL,0}$ by a LabVIEW programable driver to the MDL, a second judge and adjustment is launched. It is repeated until $|R_c(-)=R_c(+)|\leq 15 counts/s$. Such iterative procedure ensures the HOM interference maintains at the minimum coincidence, thus the two fiber paths are balanced. Through optimization, we take $\delta$=0.4 ps and $\Delta$=0.2 ps in the experiment.

\clearpage
\newpage

\section*{Acknowledgements}

This work is supported by the National Natural Science Foundation of China (Grant No. 11174282, 91336108, 61127901); the Research Equipment Development Project of the CAS, China. R.F.D. is supported by the ``Youth Talent Support Plan'' of the Organization Department, China; the Key Fund for the ``Western light'' Talent Cultivation Plan of the CAS, China. R.F.D., , S.G.Z., and T.L. are supported by the ``Cross and Cooperative'' Science and Technology Innovation Team Project of the CAS, China.

\section*{Author contributions statement}

R.F.D., S.G.Z., and T.L. conceived the experiment and supervised the overall research. R.A.Q., Y.W.Z., M.M.W., F.Y.H., S.F.W., and X.X. conducted the experiment and analysed the results.

\section*{Competing Interests}
The authors declare no competing financial interests.

\section*{Correspondence}
Correspondence and requests for materials should be addressed to Rui-Fang Dong.

\end{document}